\documentclass{elsart41}
\usepackage{graphics}
\usepackage{graphicx}
\usepackage{amssymb}
\begin{document}

\begin{frontmatter}



\title{$^{23}$Na NMR Study of Layered Superconductor Na$_x$CoO$_2\cdot y$H$_2$O}
%

\author[AA]{H.Ohta\corauthref{Name1}},
\ead{shioshio@kuchem.kyoto-u.ac.jp}
\author[AA]{C.Michioka},\,
\author[AA]{Y.Itoh},\,
\author[AA]{K.Yoshimura},\,
\author[BB]{H.Sakurai},\,
\author[BB]{E.Takayama-Muromachi},\,
\author[CC]{K.Takada},\,
\author[CC]{T.Sasaki}

\address[AA]{Department of Chemistry, Graduate School of Science, Kyoto University, Kyoto 606-8502, Japan }  
\address[BB]{Superconducting Materials Center, NIMS, Namiki, Tsukuba, Ibaraki, 305-0044, Japan }
\address[CC]{Advanced Materials Laboratry, NIMS, Namiki, Tsukuba, Ibaraki, 305-0044, Japan }

\corauth[Name1]{Corresponding author. Tel: +81-75-753-3990
fax: +81-75-753-4000}

\begin{abstract}

We measured $^{23}$Na-NMR spectra and nuclear spin-lattice relaxation rates $^{23}$(1/$T_1$) of superconducting and non-superconducting $bi$layer hydrate Na$_x$CoO$_2\cdot y$H$_2$O ( $x \sim 0.3 , y \sim 1.3$ ). The central resonance frequency shows a small but various shift due to the difference in the shielding effect by intercalated H$_2$O molecules. The different shielding effect also gives a large difference in the magnitude of $^{23}$(1/$T_1$). 

\end{abstract}

\end{frontmatter}


Recently, a number of investigations for the superconductivity in $bi$layer hydrate (BLH) sodium cobalt oxide Na$_x$CoO$_2 \cdot y$H$_2$O ( $x \sim 0.3, y \sim 1.3$ ) have been done. Different properties were reported especially in NMR studies for different samples. In a report of chemical study, it is emphasized that some Na$^+$ ions are exchanged by H$_3$O$^+$ ions during intercalation of H$_2$O molecules\cite{Takada_oxionium}. Because of the nonstoichiometric properties in BLH, i. e. ambiguities of the amount of Na$^+$, H$_2$O, and H$_3$O$^+$, similar samples are hardly obtained. 
More chemical studies are needed to clarify the origin of the magnetism and the superconductivity of BLH. 

We prepared two BLH samples by following processes\cite{Nature_Takada}. 
Powders of Co$_3$O$_4$ and Na$_2$CO$_3$ were well mixed and heated at 800 $^{\circ}$C for 20 hours in O$_2$ atmosphere. The obtained powder sample was characterized as in a single phase of Na$_{0.7}$CoO$_2$ by powder X-ray diffraction (XRD) measurements. Then Na$_{0.7}$CoO$_2$ was immersed in Br$_2$/CH$_3$CN solution for several days to deintercalate Na ions.  The dried powder was immersed in distilled water for 1 day
. The water-intercalated powder was put in a sealed chamber filled with 75 \% of relative humidity (RH). 
We prepared several samples by changing the duration kept in the chamber, and frozen at about - 10 $^{\circ}$C for quenching further change. A sample named BLH1 was kept for one week in 75 \% RH atmosphere, and BLH2 for 1 month. 

Another two samples named BLHs1 and BLHs2 were synthesized by similar process, but not at the same time with BLH1 and BLH2. BLHs1 has already been studied in \cite{Kato_sces}. BLHs2 was a derivative of BLHs1 after about 1 year in the 75 \% RH atmosphere before frozen. 

Each sample was characterized to be in a single phase of the BLH by powder XRD result. The lattice constant $c$ of BLHs1 and BLHs2 estimated by the XRD results are smaller than that of BLH1 and BLH2. This difference in $c$ indicates that the content of H$_3$O$^+$ is different between two groups. The superconducting transition temperature $T_{\mathrm{c}}$ estimated by magnetization measurements was $T_{\mathrm{c}}$ $<$ 1.8 K for BLH1 and $\sim$ 4.5 K for BLH2; $T_{\mathrm{c}}$ $\sim$ 4.7 K for BLHs1 and $\sim$ 3 K for BLHs2. 

We observed the free induction decay (FID) signals of $^{23}$Na nuclei ( $I = 3/2$ ) by NMR at 7.485 T. The Fourier transformed NMR spectra of FID signals are shown in Fig. 1. Here, $\nu_{\mathrm{ref}}$ denotes the resonance frequency of $^{23}$Na nuclei in NaCl aqueous solution at room temperature ( $\nu_{\mathrm{ref}}$ = 84.287 MHz). The satellite lines due to the nuclear quadrupole interaction may wipe out. The shift of peak resonance is very small compared with Na$_{0.7}$CoO$_2$ ( $\nu-\nu_{\mathrm{ref}} \sim$ 40 kHz )\cite{Na-NMR_allur}, and the temperature dependence of the shift is also small, suggesting a weak Co-O-Na coupling. 

The recovery curves of FID signals were measured by inversion recovery method for obtaining $^{23}$Na nuclear spin-lattice relaxation rate $^{23}$(1/$T_1$). The temperature dependence of $^{23}$(1/$T_1$) is shown in Fig. 2. The relaxation rate was estimated by the least-square fitting using the theoretical function of the central transition \mbox{( $I_z = 1/2 \leftrightarrow -1/2$ ).} While the peak shift is small and similar in each BLH as shown in Fig. 1, the absolute value of $^{23}$(1/$T_1$) is markedly different. A broad peak around 100 $\sim$ 150 K in each $^{23}$(1/$T_1$) and a rapid increase above 200 K except for BLHs2 are observed. This rapid increase in $^{23}$(1/$T_1$) may originate in the motion of Na ions\cite{Na_NMR_Gavilano}.
 
The magnitude of $^{23}$(1/$T_1$) of BLH is on two or  three order smaller than that of Na$_{0.7}$CoO$_2$\cite{Na_NMR_Gavilano,Na_NMR_Ihara}. If the hyperfine coupling of BLH is on two or three order smaller than that of Na$_{0.7}$CoO$_2$\cite{Na-NMR_allur}, the shift should be 0.05 $\sim$ 0.5 kHz.
The hyperfine coupling is thought to be shielded by the intercalated H$_2$O molecules. The difference in $^{23}$(1/$T_1$) may be attributed to the difference of compositions of H$_2$O, H$_3$O$^+$, and Na$^+$, which may cause the different surroundings around a sodium ion. Since 1/$T_1$ becomes slower with increasing the duration, the shielding effect is enhanced by the time duration in the chamber. We note that $^{59}$Co and $^{17}$O-NMR/NQR are hard to probe these difference of compositions, because Co and O nuclei are in the CoO$_2$ plane and outside of the layer consisting of the nonstoichiometric components. Therefore, $^{23}$Na-NMR should be a strong probe to elucidate the physical properties of the BLH.

In conclusion, we measured $^{23}$Na-NMR in $bi$layer hydrate sodium cobalt oxides, including nonsuperconducting $bi$layer hydrate. At the Na site, the hyperfine coupling is weakened by the intercalated H$_2$O molecules, and the nuclear spin-lattice relaxation shows a large difference in four samples with different superconducting temperatures. 

%
%
%

\begin{figure}[!ht]
\begin{center}
\includegraphics[width=0.45\textwidth]{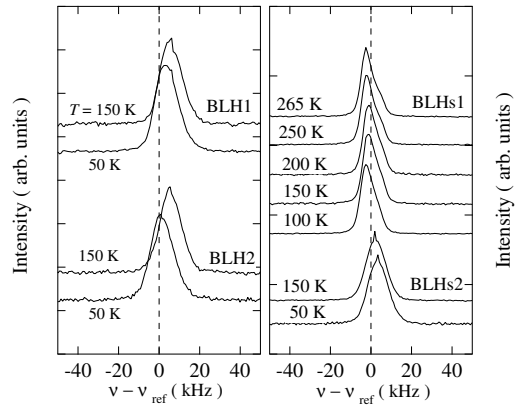}
\end{center}
\caption{$^{23}$Na-NMR spectra of BLH1,2 and BLHs1,s2. $\nu_{\mathrm{ref}}$ is the resonance frequency of $^{23}$Na in NaCl aqueous solution at room temperature ( $\nu_{\mathrm{ref}}$ = 84.287 MHz). No satellite lines were observed.}
\label{fig:spec}
\end{figure}

\begin{figure}[!ht]
\begin{center}
\includegraphics[width=0.45\textwidth]{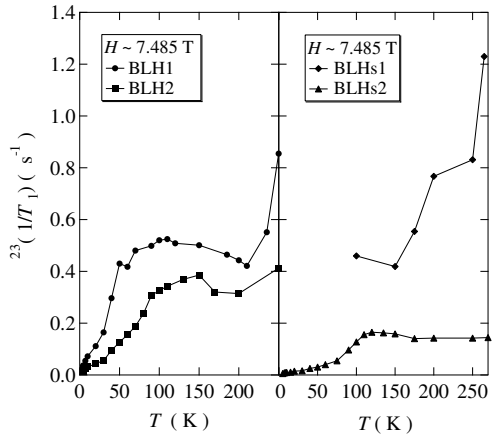}
\end{center}
\caption{Temperature dependence of $^{23}$(1/$T_1$). 
}
\label{fig:rT1}
\end{figure}

%
%
%
\section*{Acknowledgement}
This study was supported in part by Grants-in Aid for Scientific Research from Japan Society for the Promotion of Science (16076210).

\end{document}